\documentclass{aa}
\usepackage{graphicx}
\usepackage{color}
\usepackage{epstopdf}
\usepackage{txfonts}
\begin{document}

   \title{Circumstellar water vapour in M-type AGB stars:\\
   Radiative transfer models, abundances and predictions for HIFI}
\titlerunning{Circumstellar water vapour in oxygen-rich AGB stars}
 
   \author{M. Maercker
           \inst{1},
          F. L. Sch\"oier \inst{1,2},
	 H. Olofsson\inst{1,2},
	  P. Bergman\inst{2,3}
	  \and	 
	 S. Ramstedt\inst{1}
	  }
\authorrunning{M. Maercker et al.}

\offprints{M. Maercker}

   \institute{Stockholm Observatory, AlbaNova University Center, SE-106 91 Stockholm, Sweden\\
                \email{maercker@astro.su.se}
          \and Onsala Space Observatory, SE-439 92 Onsala, Sweden
          \and European Southern Observatory,  Casilla 19001, Santiago 19, Chile }

   \date{received 15 September 2007; accepted 12 December 2007}

 \abstract
{Surprisingly high amounts of $\rm{H_2O}$ have recently been reported in the circumstellar envelope around the M-type asymptotic giant branch star W~Hya. This has lead to the speculation that evaporation of icy cometary or planetary bodies might be an effective ongoing mechanism in such systems. However, substantial uncertainties remain, as the required radiative transfer modelling is difficult due to high optical depths, sub-thermal excitation and the sensitivity to the combined radiation field from the central star and  dust grains.}
{By performing a detailed radiative transfer analysis, we determine fractional abundances of circumstellar $\rm{H_2O}$ in the envelopes around six M-type asymptotic giant branch stars. The models are also used to predict $\rm{H_2O}$ spectral line emission for the upcoming Herschel/HIFI mission.}
{We use Infrared Space Observatory Long Wavelength Spectrometer spectra to constrain the circumstellar fractional abundance distribution of ortho-$\rm{H_2O}$, using a non-local thermal equilibrium, and non-local, radiative transfer code based on the accelerated lambda iteration formalism. The mass-loss rates and kinetic temperature structures for the sample stars are determined through radiative transfer modelling of CO line emission based on the Monte-Carlo method. The density and temperature profiles of the circumstellar dust grains are determined through spectral energy distribution modelling using the publicly available code Dusty.}
{The determined ortho-$\rm{H_2O}$ abundances 
lie between $2\times10^{-4}$ and $1.5\times10^{-3}$ relative to $\rm{H_2}$, with the exception of WX~Psc, which has a much lower estimated ortho-$\rm{H_2O}$ abundance of only $2\times10^{-6}$, possibly indicating $\rm{H_2O}$ adsorption onto dust grains or recent mass-loss-rate modulations. The estimated abundances are uncertain by, at best, a factor of a few.}
 {The high water abundance found for the majority of the sources suggests that either the `normal' chemical processes are very effective in producing $\rm{H_2O}$, or else non-local thermal equilibrium atmospheric chemistry, grain surface reactions, or a release of $\rm{H_2O}$ (e.g. from icy bodies like Kuiper belt objects) play a role. However, more detailed information on the physical structure and the velocity field of the region where the water vapour lines are formed is required to improve abundance estimates. We provide predictions for ortho-$\rm{H_2O}$ lines in the spectral window of Herschel/HIFI. These spectrally resolved lines cover a wide range of excitation conditions and will provide  valuable additional information on the physical and chemical properties of the inner stellar wind where $\rm{H_2O}$ is abundant.}


   \maketitle

\section{Introduction}
\label{intro}

Asymptotic giant branch stars (AGB stars) are chemically characterised according to their relative abundances of carbon and oxygen. Carbon stars have a photospheric C/O ratio $>$\,1, whereas M-type (`oxygen rich') AGB stars have C/O $<$\,1 (e.g., Russel~\cite{russel}; Beck et al.~\cite{becketal}). It is believed that M-type AGB stars evolve to carbon stars by the dredging up of nucleosynthesised carbon during the thermally pulsing AGB (TP-AGB) (e.g., Iben~\cite{iben}; Boothroyd \& Sackmann~\cite{boothroydsackmann}). The evolution along the AGB is dominated by the mass loss, starting with a low mass-loss rate $\sim$\,$10^{-8}$\,M$_{\odot}$\,yr$^{-1}$ and ending in an intense wind with rates up to $10^{-4}$\,M$_{\odot}$\,yr$^{-1}$. The photosphere and expanding circumstellar envelope (CSE) are effective producers of a variety of molecular species and microscopic dust particles, providing up to 80\% of the dust in galaxies (Whittet~\cite{whittet}). When the star reaches the end of the AGB, most of the initial mass has been lost, leaving behind only a C/O core that ionises the surrounding material and, for a short while, creates a planetary nebula (Habing~\cite{habing}). The lifetime on the AGB is estimated to be a few $\rm{\sim10^6\,yr}$ (Marigo \& Girardi~\cite{marigoco07}).

To date, about 70 circumstellar molecular species have been detected, most of them through transitions at radio wavelengths (Olofsson~\cite{olofsson07}). $\rm{H_2O}$ is of special importance in the case of M-type AGB stars, as it is possibly the dominant coolant in the inner part of the stellar wind (where it is abundant) with a large number of far-infrared (FIR) transitions (Truong-Bach et al.~\cite{truongbachetal}). $\rm{H_2O}$ line emission is also a potentially important probe of the physical and chemical properties of the inner regions of CSEs. $\rm{H_2O}$ is the main reservoir of circumstellar O, and whereas CO is abundant in all AGB stars, chemical models predict $\rm{H_2O}$ to be one of the most abundant species only in O-rich stars (e.g., Cherchneff 2006\nocite{cherchneff}). However, $\rm{H_2O}$ has also been detected in the nearby C-rich AGB star IRC+10216 (Melnick et al.~\cite{melnicketal01}; Hasegawa et al.~\cite{hasegawaetal}), but here the estimated abundance is comparatively low, and the origin of $\rm{H_2O}$ uncertain.

The atmosphere of the Earth is mostly opaque at $\rm{H_2O}$ line wavelengths and it is therefore difficult to observe $\rm{H_2O}$ from the ground, in particular in its ground vibrational state. However, Menten et al. (\cite{mentenetal}) managed to observe two vibrationally excited lines in the $\rm{\nu_2\,=\,1}$ bending mode ($\rm{5_{23}-6_{16}}$ at 336.2 GHz and $\rm{6_{61}-7_{52}}$ at 293.6 GHz) towards the red supergiant VY CMa. $\rm{H_2O}$ masers from the ground and excited vibrational states can also be detected from the ground (Menten \& Melnick~\cite{mentenmelnick}). Observations from the \emph{Infrared Space Observatory} (ISO) are not hindered by the atmosphere, and the Short- and Long-Wavelength Spectrometer (SWS and LWS) data show a rich $\rm{H_2O}$ spectrum in M-type AGB stars  ({e.g., Truong-Bach et al.~\cite{truongbachetal}; Barlow et al.~\cite{barlowetal}; Neufeld et al.~\cite{neufeldetal}; Justtanont et al.~\cite{justtanontetal}). The ortho-$\rm{H_2O}$ ground state line at 557\,GHz was also observed by the SWAS (Melnick et al.~\cite{melnicketal}) and Odin satellites (Nordh et al.~\cite{nordhetal}). In contrast to the CO radio lines, the $\rm{H_2O}$ line emission originates from closer to the star, providing information on the warm, high-density innermost layers of the CSE (Truong-Bach et al.~\cite{truongbachetal}). However, the high optical depths and the subthermal excitation of $\rm{H_2O}$ in CSEs make it a challenge to accurately model the observed $\rm{H_2O}$ spectral lines (e.g., Justtanont et al.~\cite{justtanontetal}).

We present here the results of radiative transfer modelling of the ortho-$\rm{H_2O}$ line emission in six M-type AGB stars with mass-loss rates between $\rm{1\times10^{-7}\,M_{\odot}\,yr^{-1}}$ and $\rm{4\times10^{-5}\,M_{\odot}\,yr^{-1}}$. Detailed radiative transfer models are created using the accelerated lambda iteration (ALI) method (Rybicki \& Hummer~\cite{rybickihummer91}, \cite{rybickihummer92}). The model results are fit to LWS ISO observations between 43 and 197 $\rm{\mu m}$. The mass-loss rate and kinetic temperature structure are determined through CO radio line modelling, and the dust temperature and dust density structure are determined using Dusty (Ivezi\'c et al.~\cite{ivezicetal}). The results show the ability of ALI to model circumstellar $\rm{H_2O}$ lines. We also make predictions for $\rm{H_2O}$ lines in the frequency range of the upcoming Herschel/HIFI mission. These observations will include information on the line profiles and can be used to set firmer constraints on the physical structure of the inner CSE and the $\rm{H_2O}$ abundance. Previous attempts to determine the water vapour content in stars included in our sample have been made for W~Hya (Barlow et al.~\cite{barlowetal}; Zubko \& Elitzur~\cite{zubkoelitzur}; Justtanont et al.~\cite{justtanontetal}) and R~Cas (Truong-Bach et al.~\cite{truongbachetal}). Gonz\'alez-Alfonso \& Cernicharo (\cite{gonzalezcernicharo}) used the large-velocity-gradient (LVG) method to model the 183 GHz maser line in O-rich evolved stars, to study the line intensity dependence on source properties and physical conditions. Ryde \& Eriksson (\cite{rydeeriksson}) model the $2.6-3.6\,\rm{\mu m}$ spectrum for R~Dor observed with the SWS on ISO, showing the dominance of water vapour in the spectrum.

Determining the para-$\rm{H_2O}$ abundance would be a straightforward task. However, considering the uncertainty in the abundance estimates, this would not give any firm constraints on the ortho-to-para ratio or the total $\rm{H_2O}$ abundance.

In Sect.~\ref{obs} the ISO observations are briefly described. Section~\ref{models} contains a description of the dust and CO emission line modelling. In Sect.~\ref{cslm} we describe the $\rm{H_2O}$ line model, and in Sects.~\ref{res} and~\ref{disc} we present and discuss the results, respectively. Appendix~\ref{ali} contains a description of the accelerated lambda technique, adopted in our radiative transfer model, in more detail and benchmark tests are presented. All abundances mentioned refer to the fractional abundance compared to $\rm{H_2}$, unless stated otherwise.

\section{Observations}
\label{obs}

We have used archived ISO LWS spectra ($\rm{43\,-\,197\,\mu m}$) for six M-type AGB stars (Table~\ref{isoobs}) to identify and measure the intensity of  ortho-$\rm{H_2O}$ spectral lines in the ground vibrational state. For all observations the LWS01 AOT was used (Clegg et al.~\cite{cleggetal}). The spectra were sampled at 1/4 of a resolution element, one element being $\rm{0.3\,\mu m}$ in second order ($\rm{\lambda\,\leq\,93\,\mu m}$) and $\rm{0.6\,\mu m}$ in first order ($\rm{\lambda\,\geq 80\,\mu m}$). The analysis of the data was done using the ISO Spectral Analysis Package (ISAP). The data were cleaned from cosmic rays by excluding deviant points that were only present in individual scans. All scans for each detector were median averaged. In order to get a straight (continuum-subtracted) spectrum, a baseline (typically of order 2) was fit to the data and subtracted. Finally, the ten detectors were combined to form a complete spectrum. A section of the merged spectrum for R Dor with labelled ortho- and para-$\rm{H_2O}$ transitions is shown in Fig.~\ref{isolines}. Table~\ref{lineint} gives the measured ortho-$\rm{H_2O}$ line fluxes. The observed fluxes are determined by fitting a Gaussian profile to the observed ortho-$\rm{H_2O}$ lines using available tools within ISAP. The error in the baseline fit dominates over the error in the Gaussian fit, and is $\sim$\,30\% on average (in terms of the line intensity). The fit is more difficult in the short-wavelength part of the spectrum due to the high noise levels. The measured values lie within $\pm$\,20\% of those of previous studies of $\rm{H_2O}$ emission using ISO spectra in W~Hya (Barlow et al.~\cite{barlowetal}) and R Cas (Truong-Bach et al.~\cite{truongbachetal}).

   \begin{figure*}
   \centering
   \includegraphics[angle=90,width=15cm]{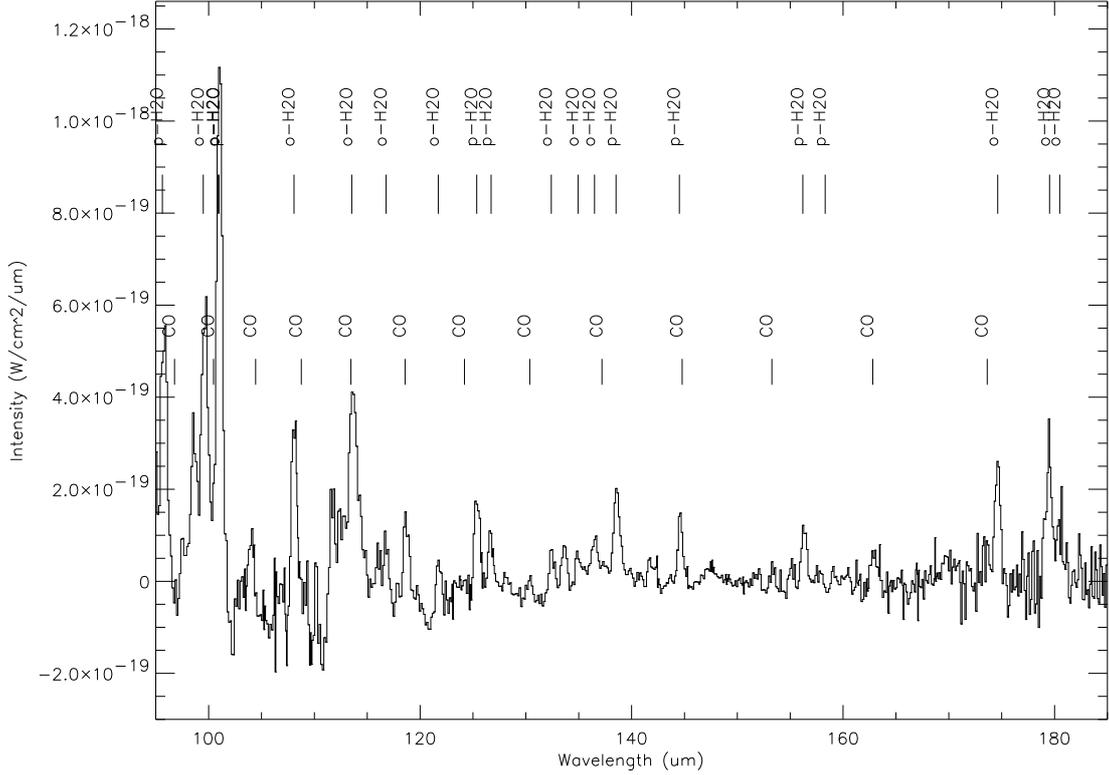}
   \caption{The ISO LWS spectrum of R Dor between 95 and 185 $\rm{\mu m}$,  including the position of ortho-$\rm{H_2O}$, para-$\rm{H_2O}$ lines, and CO lines (transitions $J\,=\,15-14$ to $J\,=\,27-26$) .}
              \label{isolines}
    \end{figure*}

\begin{table}
\caption{ISO observations for the 6 M-type AGB stars.}
\label{isoobs}
\centering
\begin{tabular}{l c c c c}
\hline\hline
Source & AOT & Observer ID & Date & Int time (sec)\\
  \hline
TX Cam & LWS01 & MBARLOW & 1997-10-10 & 3428\\
R Cas    & LWS01 & MBARLOW & 1997-06-06 & 2204\\
R Dor    & LWS01 & KERIKSSO & 1997-07-01 & 1330\\
W Hya   & LWS01 & MBARLOW & 1997-08-02 & 1778\\
WX Psc & LWS01 & MBARLOW & 1997-06-15 & 2796\\
IK Tau   & LWS01 & MBARLOW & 1997-09-02 & 3430\\
\hline
\hline
\end{tabular}
\end{table}

\begin{table*}
\caption{Measured and modelled ISO line fluxes of ortho-$\rm{H_2O}$ [in units of $\rm{10^{-20} W\,cm^{-2}]}$. Lines without fluxes were either too weak to detect or otherwise unusable, due to blending with other lines or unacceptable line fits. The dominant error in the observed fluxes is due to the baseline subtraction and is $\sim30\%$.}
\label{lineint}
\centering
\begin{tabular}{c c c c c c c c c c c c c c c c c c c}
\hline\hline
 & &\multicolumn{2}{c}{TX Cam} & &\multicolumn{2}{c}{R Cas} & &\multicolumn{2}{c}{R Dor} & &\multicolumn{2}{c}{W Hya}& &\multicolumn{2}{c}{WX
 Psc} & &\multicolumn{2}{c}{IK Tau} \\
\hline
Trans & $\lambda$ & $F\rm{_{obs}}$ & $F\rm{_{mod}}$ & &$F\rm{_{obs}}$ & $F\rm{_{mod}}$ & &$F\rm{_{obs}}$ & $F\rm{_{mod}}$ & &$F\rm{_{obs}}$ & $F\rm{_{mod}}$ & &$F\rm{_{obs}}$ & $F\rm{_{mod}}$ & &$F\rm{_{obs}}$ & $F\rm{_{mod}}$ \\
$J_{K_-K_+}-J_{K_-K_+}$& $\rm{[\mu m]}$ & & & & & & & & & & & & \\
\hline
$2_{21}-2_{12}$ & 180.49 & ---   & ---   & & --- & ---   & & 6   & 6	 &	 & ---&    & &---   & ---   & &---	    & ---   \\
$2_{12}-1_{01}$ & 179.53 & 10	 & 7	 & & 8   & 9	 & & 22  & 17	 &	 & 13 & 7  & &---   & ---   & &13	    & 10    \\
$3_{03}-2_{12}$ & 174.63 & 10	 & 6	 & & 7   & 8	 & & 17  & 20	 &	 & 12 & 8  & &---   & ---   & &15	    & 8     \\
$3_{30}-3_{21}$ & 136.50 & ---   & ---   & & 3   & 3	 & & 8   & 8	 &	 & 3  & 4  & &1     & 1     & &6	    & 7     \\
$5_{14}-5_{05}$ & 134.94 & ---   & ---   & & 2   & 2	 & & 6   & 6	 &	 & 3  & 4  & &---   & ---   & &5	    & 5     \\
$4_{23}-4_{14}$ & 132.41 & ---   & ---   & & 4   & 3	 & & 7   & 9	 &	 & 4  & 4  & &2     & 1     & &5	    & 6     \\
$4_{32}-4_{23}$ & 121.72 & ---   & ---   & & 4   & 4	 & & 9   & 9	 &	 & 6  & 5  & &1     & 1     & &7	    & 10    \\
$7_{34}-6_{43}$ & 116.78 & ---   & ---   & & --- & ---   & & 10  & 8	 &	 & 9  & 7  & &---   & ---   & &---	    & ---   \\
$4_{14}-3_{03}$ & 113.54 & 9	 & 8	 & & 14  & 10	 & & --- & ---   &	 & 17 & 11 & &---   & ---   & &20	    & 15    \\
$2_{21}-1_{10}$ & 108.07 & 11	 & 11	 & & 16  & 14	 & & 31  & 31	 &	 & 18 & 15 & &---   & ---   & &21	    & 19    \\
$5_{05}-4_{14}$ & 99.49  & 10	 & 10	 & & 16  & 17	 & & 52  & 39	 &	 & 20 & 16 & &---   & ---   & &50	    & 19    \\
$6_{16}-5_{05}$ & 82.03  & ---   & ---   & & --- & ---   & & 58  & 40	 &	 & ---& ---& &---   & ---   & &41	    & 26    \\
$4_{23}-3_{12}$ & 78.74  & 14	 & 13	 & & 25  & 17	 & & --- & ---   &	 & ---& ---& &---   & ---   & &57	    & 33    \\
$3_{21}-2_{12}$ & 75.38  & 23	 & 23	 & & 28  & 28	 & & 43  & 63	 &	 & 35 & 31 & &---   & ---   & &50	    & 44    \\
$7_{07}-6_{16}$ & 71.95  & ---   & ---   & & --- & ---   & & --- & ---   &	 & 24 & 23 & &---   & ---   & &37	    & 39    \\
$3_{30}-2_{21}$ & 66.44  & 18	 & 22	 & & 35  & 25	 & & 59  & 65	 &	 & 26 & 31 & &9     & 12    & &48	    & 54    \\
$5_{32}-4_{23}$ & 47.97  & ---   & ---   & & --- & ---   & & 109 & 89	 &	 & 52 & 46 & &---   & ---   & &105	    & 132   \\
$5_{23}-4_{14}$ & 45.11  & ---   & ---   & & --- & ---   & & 155 &102	 &	 & 70 & 64 & &---   & ---   & &161	    & 161   \\

\hline
\hline
\end{tabular}
\end{table*}

\section{Circumstellar dust and CO line  modelling}
\label{models}
\subsection{Dust emission modelling using Dusty}
\label{dusty}
The dust optical depth and dust temperature profile were determined using the continuum radiative transfer code Dusty (Ivezi\'c et al.~\cite{ivezicetal}). The observational constraints are in the form of spectral energy distributions (SEDs) and consist of 2MASS and IRAS fluxes. A `standard model' containing amorphous silicate dust was used, with optical constants adopted from Justtanont \& Tielens (\cite{justtanonttielens}). The standard model assumes the mass loss to be spherically symmetric with a constant rate and expansion velocity, resulting in a density profile with $\rho \propto r^{-2}$. Prompt dust formation is assumed at the condensation temperature $T_c$, which is treated as a free parameter. Additionally, the dust grains are assumed to be spherical with a radius of $\rm{0.1\,\mu m}$. The dust optical depth and dust condensation temperature were adjusted in order to find the best fit to the SED using a $\chi^2$-statistic. A more detailed description of the modelling procedure is given in Sch\"oier et al. (\cite{schoieretal}).

The luminosities of the Miras are derived using a period-luminosity (PL) relationship  based on bolometric magnitudes (Feast et al.~\cite{feastetal}). The distances are obtained from the apparent bolometric magnitude obtained in the SED modelling. For semi-regular objects, Knapp et al. (\cite{knappetal}) give a PL relationship based on K-band magnitudes. Since this only uses information in one point of the SED (the K-band magnitude), we instead used the revised Hipparcos distances for W~Hya and R~Dor (Knapp et al.~\cite{knappetal}), and determined the luminosity through the SED modelling. 

 \subsection{Modelling of CO lines}
 \label{COlinemod}
The CO line modelling is based on the Monte-Carlo method presented in detail in Sch\"oier \& Olofsson (\cite{schoierolofsson}) and Olofsson et al. (\cite{olofssonetal}). The radiative transfer model includes the lowest 40 rotational levels in the ground and first vibrational states. The collision rates are taken from Flower (\cite{flower}) for CO-$\rm{H_2}$, and they are extrapolated to higher rotational temperatures as was done in Sch\"oier et al. (\cite{schoieretal05}). The collisional rates between CO and ortho- and para-$\rm{H_2}$ were weighted assuming an ortho- to para-$\rm{H_2}$ ratio of 3. A CO fractional abundance ($\rm{CO/H_2}$) of $2\times10^{-4}$ is adopted, a value typically used for O-rich AGB stars. The model includes cooling terms from the adiabatic expansion of the envelope and radiative line cooling through CO and $\rm{H_2}$. The dominant heating term is given by dust-gas collisions. The density distribution and temperature structure of the dust grains are taken from the dust radiative transfer modelling (see Sect.~\ref{dusty}), while the parameters for grain size, grain mass density and dust-to-gas ratio are combined into one parameter \emph{h}. The \emph{h}-parameter is, in principal, a free parameter in the CO modelling, and is chosen to be consistent with previous models (Olofsson et al.~\cite{olofssonetal}: W~Hya and R~Dor; Gonz\'alez Delgado et al.~\cite{delgadoetal}: R~Cas; Ramstedt et al.~\cite{ramstedtetal07}: TX~Cam, WX~Psc, and IK~Tau). The outer radius is set by the photodissociation of CO (Stanek et al.~\cite{staneketal}, based on results by Mamon et al.~\cite{mamonetal}). The expansion velocity and micro-turbulent velocity ($\rm{0.5\,km\,s^{-1}}$ doppler width), with local thermal (Gaussian) line broadening, are also set in the model. This leaves the mass-loss rate as the remaining free parameter that affects the model line fluxes. The best fit model is then determined by fitting the model line fluxes to the observed fluxes and minimising  $\chi^2$, where
\begin{equation}
\label{chi2}
\chi^2=\sum{{(I_{\rm{obs,i}}-I_{\rm{mod,i}})^2} \over {\sigma_{\rm{i}}^2}}
\end{equation}
and $I_{\rm{obs}}$ and $I_{\rm{mod}}$ are the observed and modelled line fluxes, respectively, $\sigma_{\rm{i}}$ the uncertainty in observation $i$, and the sum goes over all lines included.
 
The line profile is not taken into consideration, but we note that in the case of W~Hya, Justtanont et al. (\cite{justtanontetal})  had to reduce the CO photodissociation radius and increase the mass-loss rate in order to fit the line profile. However, there is sufficient uncertainty in the underlying assumptions of the circumstellar model that we prefer to treat all objects in the same way and avoid individual adjustments of, e.g., the CO envelope size. 

The uncertainty in the mass-loss rate is estimated to be $\rm{\pm\,50\%}$ within the adopted model (Sch\"oier \& Olofsson~\cite{schoierolofsson}). Previous mass-loss-rate estimates by Gonz\'alez Delgado et al.~(\cite{delgadoetal}) and Olofsson et al.~(\cite{olofssonetal}) differ somewhat from the results in Table~\ref{modres}. The difference is less than 10\% for the low-mass-loss-rate and intermediate-mass-loss-rate objects, but more for the high-mass-loss-rate objects WX Psc and IK Tau ($\rm{\pm\,50\%}$). This is due to the combination of new adopted distances, inclusion of new data, updated collisional rate coefficients and the inclusion of thermal radiation from dust grains in our CO models. The observed CO line fluxes, used in the model fitting, are presented in Table~\ref{coobs}. IK~Tau, WX~Psc and TX~Cam were modelled by Ramstedt et al. (\cite{ramstedtetal07}). The parameters derived from the dust and CO modelling are presented in Table~\ref{modres}.

\begin{table}
\caption{Observed CO line fluxes in main beam brightness scale. Observations were taken at Onsala Space Observatory, JCMT, and SEST (see references for details).}
\label{coobs}
\centering
\begin{tabular}{l c c c c}
\hline\hline
Source 	& $J$\,$=$\,1\,$\rightarrow$\,0 		& $J$\,$=$\,2\,$\rightarrow$\,1 		& $J$\,$=$\,3\,$\rightarrow$\,2 		& $J$\,$=$\,4\,$\rightarrow$\,3  \\
 		&[$\rm{K\,km\,s^{-1}}$]&[$\rm{K\,km\,s^{-1}}$]&[$\rm{K\,km\,s^{-1}}$]&[$\rm{K\,km\,s^{-1}}$]\\
  \hline
TX Cam	& 20.1$^e${\phantom{3}}	            		& 60.6$^e$	& {\phantom{1}}71.1$^e$	& 148.7$^e$			\\
R Cas	& {\phantom{3}}8.4$^a${\phantom{3}}  	& 32.1$^b$	& 100.2$^b$	             	& {\phantom{1}}89.6$^b$	\\
R Dor	& {\phantom{3}}5.0$^a${\phantom{3}}  	& 41.6$^a$	& {\phantom{1}}70.4$^a$	& -					\\
W Hya	& {\phantom{3}}0.83$^a$	            		& 18.7$^c$	& {\phantom{1}}40.2$^d$	& {\phantom{1}}44.6$^c$	\\
WX Psc 	& 44.4$^e${\phantom{3}}	            		& 80.6$^e$	& {\phantom{1}}70.1$^e$	& {\phantom{1}}83.8$^e$	\\
IK Tau 	&46.2$^e${\phantom{3}}	            		& 95.1$^e$	& 124.5$^e$	           	& 129.6$^e$			 \\
\hline
\hline
\end{tabular}
\begin{list}{}{}
\item[$^{\rm{a}}$]Kerschbaum \& Olofsson (\cite{kerschbaumolofsson}), $^{\rm{b}}$ Delgado et al. (\cite{delgadoetal}), $^{\rm{c}}$ JCMT archive, $^{\rm{d}}$ Justtanont et al. (\cite{justtanontetal}), $^{\rm{e}}$ Ramstedt et al. (\cite{ramstedtetal07})
\end{list}

\end{table}

 \section{Circumstellar $\rm{H_2O}$ line modelling (ALI)}
\label{cslm}
We have developed a detailed radiative transfer code based on the accelerated lambda iteration method, in order to accurately model circumstellar $\rm{H_2O}$ line emission (Bergman, internal report). The Monte Carlo method is not well suited for such problems due to its slow convergence at high optical depths. The ALI code was previously used in Justtanont et al. (\cite{justtanontetal}), but a number of modifications have subsequently been made to be able to treat the inclusion of dust emission and vibrationally excited energy levels. The accelerated lambda technique and its numerical implementation is described in more detail in appendix~\ref{ali}, including results of benchmarking tests against the standard problems presented in van Zadelhoff et al. (\cite{zadelhoffetal}).

\subsection{Dependence on numerical parameters}
\label{numpar}
ALI calculates the mean intensity at each radial point in the CSE, including the contribution from all other points in the envelope. This is done by paying special attention to a proper sampling of the central source. Typically, 50 radial points (distributed in a logarithmic way) and 32 angular points were used. 
The dependence on the radial and angular sampling was tested for the low mass-loss-rate object R~Dor and intermediate mass-loss-rate object TX~Cam by increasing the two parameters by a factor of two. The differences in the modelled line fluxes were small in both cases, less than 2\%.

After each iteration the new level populations are calculated, using a user-defined fraction of the old level populations (here set to 0.8) to prevent the models from diverging too easily. The models are said to converge when the average change in the level populations between iterations is less than a defined limit. This limit was also tested by running models with different limits for R Dor and IK Tau. For an average change between iterations in the level populations of less than $10^{-3}$ the models changed by less than 0.1\% This limit was subsequently used for all models.

\subsection{Modelling of $\rm{H_2O}$ lines}
\label{modinp}
The radiative transfer model includes the 45 lowest levels in the ground and first excited vibrational states (the bending mode $\rm{\nu_2}=1$, $\rm{6.3\,\mu m}$ or $\approx$\,2300\,K above the ground state) of ortho-$\rm{H_2O}$, while excitation through the stretching modes can be neglected. For the high mass-loss-rate objects IK Tau and WX Psc, only the ground vibrational state was included in order to improve the convergence. The effect of excluding the $\rm{\nu_2}$ state on the resulting model line fluxes was tested and found to be insignificant ($\leq$\,4\%) for these high-mass-loss-rate objects (see Sect.~\ref{paramdep}). The collisional rates in the ground state are  taken from the rates of $\rm{H_2O}$ with $\rm{He}$, corrected by a factor of 1.4 to account for collisions with $\rm{H_2}$ (Green et al.~\cite{greenetal}). Phillips et al. (\cite{phillipsetal}) suggest that this may lead to rather different rates compared to those obtained when including $\rm{H_2}$ in the calculation. However, they only calculate collisional rates between $\rm{H_2O}$ and $\rm{H_2}$ up to a temperature of 140\,K, while typical temperatures in the CSEs of AGB stars can be up to 1000\,K. The importance of collisions within the $\rm{\nu_2}$ state and between the ground and $\rm{\nu_2}$ state was also tested (see Sect.~\ref{paramdep}). Rotational collision rates within the first excited state are taken to be the same as for the ground state, while collisions between the ground and excited $\rm{\nu_2}$ state are based on the ground state rotational collision rate coefficients scaled by a factor of 0.01. The molecular data for $\rm{H_2O}$ was taken from the Leiden Atomic and Molecular Database\footnote{available at http://www.strw.leidenuniv.nl/$\sim$moldata} (LAMDA) (Sch\"oier et al.~\cite{schoieretal05}).

As with the CO models, the density distribution and temperature structure of the dust grains are taken from the dust radiative transfer modelling (see Sect.~\ref{dusty}). From the CO line modelling the mass-loss rate, expansion velocity, and kinetic temperature profile are included (see Sect.~\ref{COlinemod} and Table~\ref{modres}). The inner radius of the CSE is set to the inner radius of the dust envelope, i.e. the dust condensation radius. Setting a smaller radius does not affect the line fluxes. For all models, a Gaussian distribution of circumstellar $\rm{H_2O}$ is assumed. The maximum radius is defined by the e-folding radius $r\rm{_e}$ (i.e., where the abundance has decreased to 37\%) and is set to 3\,$\times$\,$r\rm{_e}$. By varying $r\rm{_e}$, the size of the circumstellar $\rm{H_2O}$ envelope is treated as a free parameter. 

An estimate of $r\rm{_e}$ is given by the results of theoretical models of $\rm{H_2O}$ photodissociation (Netzer \& Knapp~\cite{netzerknapp}), where the radius at which the abundance of OH (the photodissociation product of $\rm{H_2O}$) is highest is given by

\begin{equation}
\label{outrad}
r_{\rm{NK}} = 5.4\times10^{16}  \left(\frac{\dot{M}}{10^{-5}\,\rm{M_{\odot}\,yr^{-1}}}\right) ^{0.7}\left(\frac{v_{\rm{exp}}}{\rm{km\,s^{-1}}}\right)^{-0.4}\,\rm{cm.}
\end{equation}

The remaining parameter to fit is the ortho-$\rm{H_2O}$ abundance $f\rm{_{o-H_2O}}$ relative to $\rm{H_2}$. The model lines are fit to the line fluxes of the observed spectra in the same way as for the CO emission line models. The best fit model was determined, as for the CO modelling, by minimising $\chi^2$ as defined in Eq.~\ref{chi2}, with $r_{\rm{e}}$ and $f_{\rm{o-H_2O}}$ as free parameters.

\begin{figure}
   \centering
   \includegraphics[]{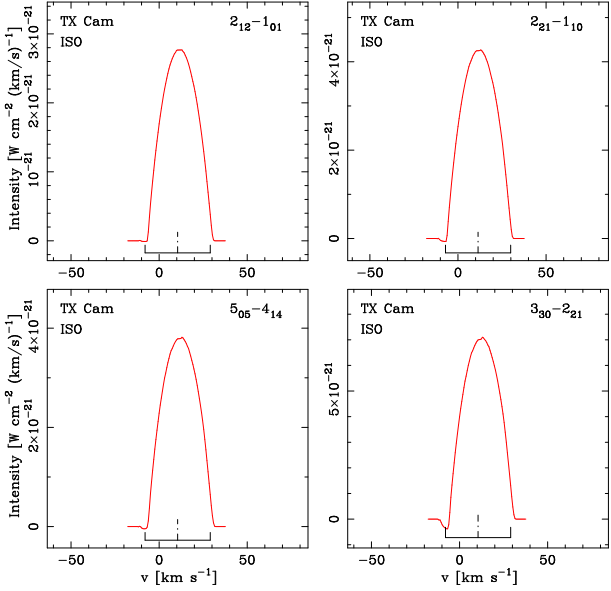}
   \caption{Modelled ISO ortho-$\rm{H_2O}$ lines for TX Cam. The dot-dashed line marks the stellar $v_{\rm{LSR}}$, the solid line shows  $v_{\rm{LSR}}\pm v_{\rm{exp}}$. Self-absorption and weak P-Cygni profiles are present in the model line profiles.}
              \label{isomod}
    \end{figure}

 \section{Results}
 \label{res}
 \subsection{Modelled lines and ortho-$H_2O$ abundances}
 \label{abundances}
For each object a grid, of models with varying ortho-$\rm{H_2O}$ abundance and $\rm{H_2O}$ envelope size was calculated. The resulting $\chi^2$ maps are shown in Fig.~\ref{gridmaps}. Between 4 and 16 lines are matched for each object. The fluxes of the best-fit modelled ortho-$\rm{H_2O}$ lines are shown in Table~\ref{lineint}, together with the observed fluxes. For the low-mass-loss-rate objects the abundance and $r_{\rm{e}}$ can be reasonably well determined, while setting constraints for $r_{\rm{e}}$ becomes increasingly difficult with increasing mass-loss-rate. In the high-mass-loss-rate objects, the emitting region is excitation limited. Consequently, the emission is not very sensitive to the envelope size. As a result, it is difficult to constrain the size of the of the $\rm{H_2O}$ envelope for these objects. For the low-mass-loss-rate objects, the excitation can be high throughout the envelope, thus making the emission sensitive to the envelope size, i.e. the emission can be said to be photodissociation limited. The grids are terminated in envelope size when an increase in $r_{\rm{e}}$ does not lead to a significant change in line fluxes anymore, or when the 1$\sigma$ contour closes. For IK~Tau and WX~Psc a lower limit could not be determined, as smaller envelope sizes require a higher abundance, for which the models could not be made to converge. The $\chi^2$ maps are truncated at $\approx0.1\times\,r\rm{_{NK}}$ for these objects. In the cases where an outer radius could be determined, the radius given by Eq.~\ref{outrad} lies within the $1\,\sigma$ contour of the lowest $\chi^2$ model radius. Table~\ref{modres} gives the radius from the theoretical model ($r_{\rm{NK}}$), the radius of the lowest $\chi^2$ model ($r_{\rm{e}}$), and the corresponding abundance estimates ($f\rm{_{NK}}$ and $f\rm{_e}$, respectively). From here on, the `best-fit' models refer to the results obtained using $r\rm{_{NK}}$ from Eq.~\ref{outrad}.

\begin{figure*}
   \centering
\includegraphics[]{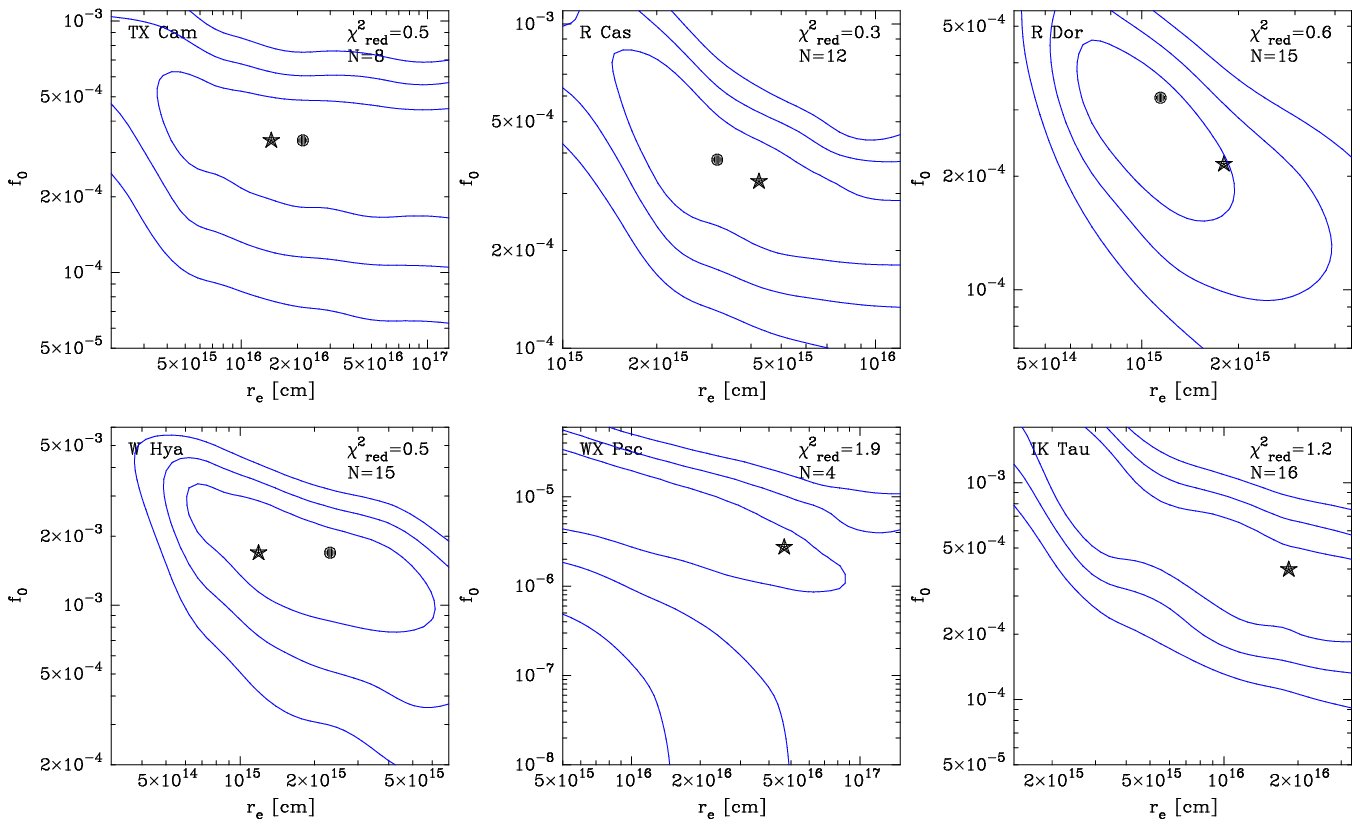}
   \caption{$\chi^2$ maps in ortho-$\rm{H_2O}$ abundance and radius of the $\rm{H_2O}$ envelope for all stars in the sample. The contours give the 1, 2 and 3$\sigma$ limits. The stars mark the best model based on the theoretical radius given by Eq.~\ref{outrad}, the dots mark the models with the lowest $\chi^2$ values. The $\chi^2_{\mathrm{red}}$ values in the figures refer to the best model fits.}
              \label{gridmaps}
    \end{figure*}

The resulting abundances of circumstellar ortho-$\rm{H_2O}$, from fitting the model lines to the observed ISO lines, lie between 2\,$\times$\,$10^{-6}$ and 1.5\,$\times$\,$10^{-3}$ relative to $\rm{H_2}$ (Table~\ref{modres}). Four of the modelled ISO lines of TX~Cam are shown in Fig.~\ref{isomod} as an example of an intermediate-mass-loss-rate object. Self absorption is present in all the lines, whereas weak P-Cygni profiles are seen in the higher-frequency lines, where the continuum emission from thermal dust grains is stronger. Since the line shapes are not spectrally resolved in the ISO spectra, only the fluxes were fit. HIFI will be able to resolve the line profiles, and these can then be taken into consideration when fitting the models (Sect.~\ref{hifi}). 
    
Figure~\ref{ex} shows the kinetic temperature profile compared to the excitation temperature for four lines for W~Hya and IK~Tau (upper panels), and the tangential optical depth at line centre for the same lines as a function of distance from the central star (lower panels). It should be pointed out that the optical depth in the line wings usually is considerably higher. Two of the lines are observed by ISO, while three of the lines will be observable by HIFI. In general, the lines are subthermally excited, and the $J_{K_-K_+}-J_{K_-K_+}=5_{32}-4_{41}$ transition at 620.9 GHz shows maser action. The lines are more thermalised in the high-mass-loss-rate object IK~Tau. The tangential optical depth shows the regions where the lines are created, and this indicates that the ISO observations do indeed probe the inner regions of the $\rm{H_2O}$ envelope. W~Hya has a smaller envelope due to the lower mass-loss-rate and consequently the lines are formed closer to the star.

\begin{table*}
\caption{Results of the dust emission, and CO and $\rm{H_2O}$ line modelling (see text for details). $r\rm{_{NK}}$ and $r\rm{_e}$ are  the theoretical model radius and the e-folding radius, respectively. The corresponding $\rm{H_2O}$ abundances for both radii are given under $f\rm{_NK}$ and $f\rm{_e}$, respectively. The $\chi_{red}^2$ values refer to the best $\rm{H_2O}$ model fits using $r\rm{_{NK}}$. Likewise, diff gives the average difference of the modelled $\rm{H_2O}$ line fluxes using $r\rm{_{NK}}$ to the observations in \%.}
\label{modres}
\centering
\begin{tabular}{l l c c c c c c c c c c c c}
\hline\hline
Source	& Type &$L\rm{^a}$ 	& $D\rm{^b}$   & $\rm{\tau_{dust}}$ & \emph{\.M} & $v\rm{_{exp}}$ & $r\rm{_{NK}}$ & $r\rm{_e}$ &$f\rm{_{NK}}$ & $f\rm{_e}$ & $\rm{\chi^2_{red}}$ & diff	\\
	& & [$\rm{L_{\odot}}$] & [pc] & 10 $\rm{\mu m}$ & [$10^{-6}\,M_{\odot}\,yr^{-1}$] & [$\rm{km s^{-1}}$]  & [$\rm{10^{15}cm}$] &[$\rm{10^{15}cm}$] & [$10^{-4}$] &[$10^{-4}$] & & \% \\
  \hline
TX Cam	& M & 11900	             & 440 	          & 0.4{\phantom{5}} & {\phantom{5}}7.0 & 18.5  	         & 13.1             & 20.0             &  {\phantom{1}}3.0{\phantom{2}} & {\phantom{1}}3.0 & 0.6   & {\phantom{1}}-6.8    \\
R Cas	& M & {\phantom{1}}8725	 & 172	           & 0.05	        & {\phantom{5}}0.9 & 10.5	         & {\phantom{5}}3.9 & {\phantom{2}}3.0 &  {\phantom{1}}3.0{\phantom{2}} & {\phantom{1}}3.5 & 0.3   & {\phantom{1}}-4.8   \\
R Dor	& SRb & {\phantom{1}}6500 & {\phantom{1}}59 & 0.03	        & {\phantom{5}}0.2 & {\phantom{1}}6.0 & {\phantom{5}}1.7 & {\phantom{2}}1.1 &  {\phantom{1}}2.0{\phantom{2}} & {\phantom{1}}3.0 & 0.7   & {\phantom{1}}-1.9   \\
W Hya	& SRa & {\phantom{1}}5400 & {\phantom{1}}78 & 0.07	        & {\phantom{5}}0.1 & {\phantom{1}}7.2	 & {\phantom{5}}1.1 & {\phantom{2}}2.2 &  15.0{\phantom{2}}             & 15.0             & 0.6   & -10.4   \\
WX Psc 	& M & 14600	             & 720 	          & 3.0{\phantom{5}} & 40{\phantom{.0}} & 19.3  	         & 43.0             & -                & {\phantom{1}}0.02              & -                & 2.7   & {\phantom{1}}-7.3  \\
IK Tau 	& M & 10400	             & 300 	          & 1.2{\phantom{5}}	& 10{\phantom{.0}} & 19.0  	         & 17.0             & -                & {\phantom{1}}3.5{\phantom{2}}  & -                & 1.2   & {\phantom{1}}-8.5  \\
\hline
\hline
\end{tabular}
\begin{list}{}{}
\item[$^{\rm{a}}$]From period-luminosity relation (Feast et al.~\cite{feastetal}), except for R~Dor and W~Hya, where the Hipparcos distances and the dust modelling are used.
\item[$^{\rm{b}}$]Determined in the dust modelling using $L$, except for R~Dor and W~Hya, where the Hipparcos distances are used (Knapp et al. \cite{knappetal}).
\end{list}
\end{table*}

\subsection{Constraining the $\rm{H_2O}$ abundance distribution}
\label{constdist}

Several test models were calculated in order to see how the results depend on the ortho-$\rm{H_2O}$ abundance, mass-loss-rate, and size of the circumstellar envelope. This was done by varying one of the parameters, while all other parameters were held constant. Table~\ref{radtest} shows the results in the cases of R~Dor and TX~Cam, as examples of low and intermediate-mass-loss-rate objects. 

Increasing and decreasing the abundance by a factor of two changes the line fluxes on average by $\pm20\%$ and $\pm10\%$ for the low- and high-mass-loss-rate objects, respectively (i.e. less than the observational error in the ISO lines). In general, the error in the $\rm{H_2O}$ abundances is estimated to be of the order $\pm50\%$ within the adopted model.

Varying the mass-loss rate by $\pm50\%$ changes the line fluxes by $\approx$\,$\pm$\,30\%. The change in mass-loss rate affects the density distribution, and hence the efficiency of the collisional excitation, making the lines more sensitive for changes in the mass-loss rate than for changes in the abundance.

Increasing and decreasing $r_{\rm{e}}$ by a factor of two changes the line fluxes by less than $\pm20\%$. The emitting region is excitation limited, instead of limited by the size of the $\rm{H_2O}$ envelope, making it difficult to provide good constraints on $r_{\rm{e}}$. The position of OH masers for W~Hya ($0.8-1.4\times10^{15}$\,cm from the central star, see Szymczak et al.~(\cite{szymczaketal})) and for R~Cas ($1.3-2.6\times10^{15}$\,cm from the central star, see Chapman et al.~(\cite{chapmanetal})), indicate that the $\rm{H_2O}$ envelope might actually be smaller than given by Eq.~\ref{outrad}. 

We conclude that the saturation of the lines, and hence the insensitivity of the integrated intensity to the abundance, is the major source of uncertainty in the determined abundances. Taking into account the uncertainty in the mass-loss rate (Ramstedt et al.~\cite{ramstedtetal07}), the line fluxes, and in the model-dependent assumptions (such as the molecular distribution and velocity field), the determined abundances are accurate to within a factor of approximately 5.

  \begin{figure*}
   \centering
   \includegraphics[width=15cm]{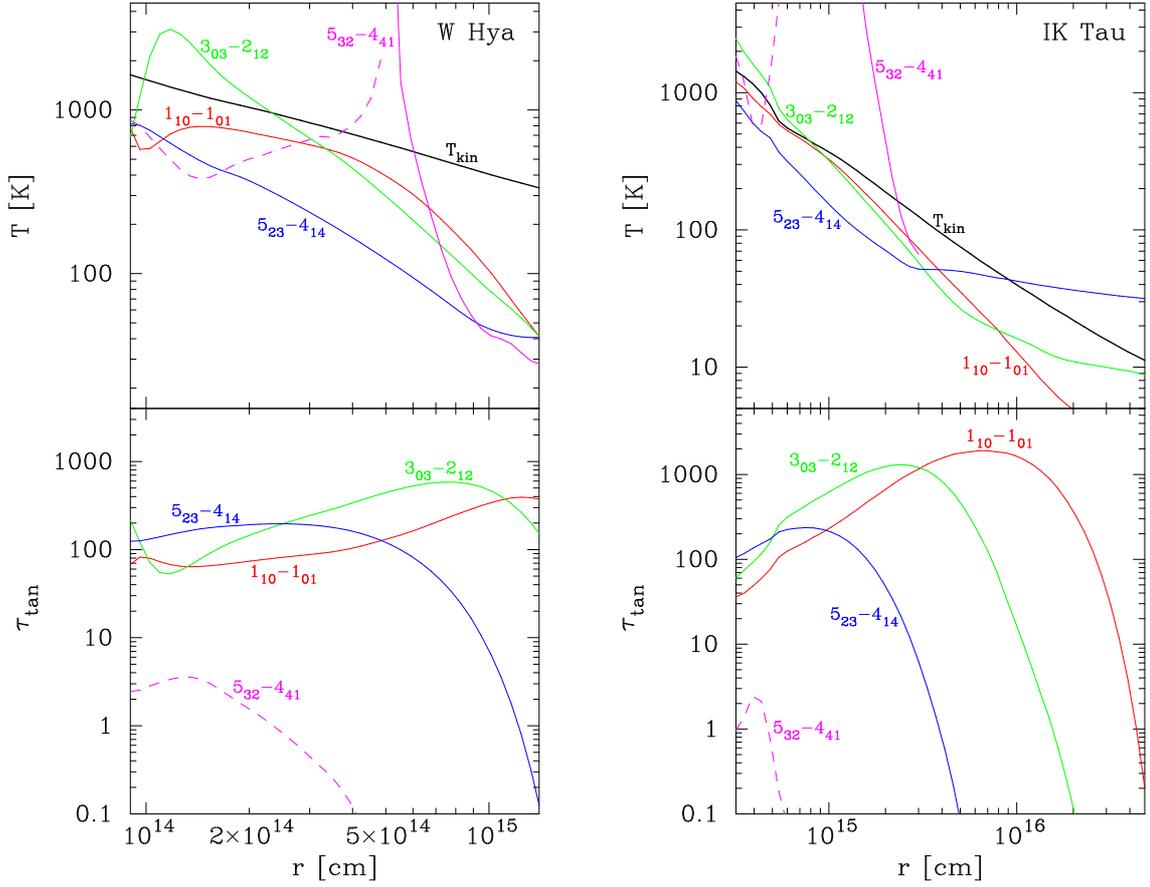}
   \caption{Kinetic temperature profile and excitation temperatures for four modelled lines (indicated by their $J_{K_-K_+}-J_{K_-K_+}$ quantum numbers) in W~Hya and IK~Tau (upper panels), and the tangential optical depths (at line centre) for the same lines (lower panels). The lines include a high transition line observed with ISO ($5_{23}-4_{14}$ at 45 $\rm{\mu m}$), a low lying transition observed with ISO ($3_{03}-2_{12}$ at 174.6 $\rm{\mu m}$), and the 557 GHz line ($1_{10}-1_{01}$) observed by Odin and SWAS (the latter two can also be observed by HIFI). A line, exhibiting maser action, which can be observed by HIFI is shown ($5_{32}-4_{41}$ at 621 GHz); the dashed line indicates the region where the level populations are inverted, and shows the absolute value of the optical depth in the lower panel. For radii larger than $3\times10^{15}$cm, numerical noise dominates the $5_{32}-4_{41}$ transition in IK~Tau and the model is consequently terminated at this radius.}
              \label{ex}
    \end{figure*}

\subsection{Dependences on `fixed' parameters}
\label{paramdep}
Additional test models were calculated to examine the dependence of the line fluxes on the `fixed' parameters (see Table~\ref{radtest}). As before, only one parameter was varied, while all other parameters were held fixed. The effect of excitation to the first excited vibrational state was tested for the four stars for which models including this state converged. Excluding excitation to this state leads to a stronger decrease of the line fluxes for the low-mass-loss rate objects than for the intermediate and high-mass-loss rate objects. Hence the error introduced by not including the first excited vibrational state for IK~Tau and WX~Psc is expected to be well below the uncertainty due to the observed line fluxes. For the low-mass-loss-rate objects, the inclusion of vibrational excitation results in significantly lower abundances compared to models that do not include the excited vibrational state (Justtanont et al.~\cite{justtanontetal}). The variation of the line fluxes when excluding the dust radiation field and the central star was also examined in test models for the low-mass-loss-rate object R~Dor and intermediate-mass-loss-rate object TX~Cam (Table~\ref{radtest}). It must be kept in mind that the changes in line fluxes all lie within (or close to) the observational uncertainties. This is especially true for TX~Cam, where the change between different models is at most $12\%$ on average.

Including the rotational collision rates within the $\rm{\nu_2}=1$ state and between the ground state and first excited state, as described in Sect.~\ref{modinp}, does not change the fluxes significantly for any of the stars, showing that the dust radiation field and radiation from the central star are more important for the excitation of the $\rm{\nu_2}=1$ state than collisions. The role of collisions within the vibrational ground state is tested by setting all collisional coefficients to zero, decreasing the observed line fluxes, on average, by 17\% in the case of R~Dor and 58\% for TX~Cam. As the lines are generally subthermally excited, the error in the collision rates in the ground state is likely to affect the modelled line fluxes (see Sect.~\ref{modinp}). 

Since the ISO $\rm{H_2O}$ emission lines probe the inner parts of the CSE, the expansion velocity in the region where the lines are created may not have reached its terminal value as determined in the CO line modelling (see Sect.~\ref{COlinemod}). [Note that changing the expansion velocity, while keeping the mass-loss-rate constant, effectively changes the density profile.] Decreasing the expansion velocity by 20\% only changes the line fluxes by +6\% for both TX~Cam and R~Dor, showing that the uncertainty in the expansion velocity does not contribute significantly to the error in the estimated abundances.

\begin{table}
\caption{Parameter dependence of the model ISO line fluxes for the low-mass-loss-rate object R~Dor and the intermediate-mass-loss-rate object TX~Cam. Column 2 gives the change of the various parameters relative to their best-fit value (Table~\ref{modres}). Columns 3 and 4 give the average difference of all model line fluxes compared to the best-fit model.}
\label{radtest}
\centering
\begin{tabular}{l c c c}
\hline\hline
param. & change &R Dor & TX Cam \\
\hline
$f_{\rm{e}}$     & +100\%	          & +20\%	         & +10\%	\\
				& {\phantom{1}}--50\% & --20\%	         & --10\%	\\
$R_{\rm{mod}}$	& +100\%              & +20\%	         & +20\%	\\
				& {\phantom{1}}--50\% & --20\%	         & --20\%	\\
$\emph{\.M}$		& {\phantom{1}}+50\%	 & +30\%	             & +30\%	\\
				& {\phantom{1}}--50\% & --30\%	         & --30\%	\\
$v_{\rm{exp}}$	& {\phantom{1}}--20\% & {\phantom{1}}+6\% & {\phantom{1}}+6\%	\\
$T_{\rm{kin}}$	& {\phantom{1}}$\pm20\%$ &$\pm15\%$       &$\pm15\%$\\
collisions		& excl.	              & --17\%	         & --58\%	\\
$\nu_2$			& excl.	              & --35\%	         & {\phantom{1}}--4\%	\\
dust				& excl.	              & --11\%	         & --12\%	\\
star				& excl.	              & --19\%	         & {\phantom{1}}--3\%	\\
\hline
\end{tabular}
\end{table}

\subsection{$H_2O$ line cooling}
\label{cooling}
In addition to CO and $\rm{H_2}$ cooling by line emission, $\rm{H_2O}$ could also be an important coolant in the inner part of CSEs where it is abundant. From the excitation analysis, the cooling (or heating) arising from line emission can be estimated using
\begin{equation}
\Lambda =n_{\rm{H_2}} \sum_l \sum_{u>l}(n_l\gamma_{lu}-n_u\gamma_{ul})h\nu_{ul},
\end{equation}
where $n_u$ and $n_l$ are the level populations in the upper and lower levels participating in the transition (with rest frequency $\nu_{ul}$) and $\gamma_{ul}$ and $\gamma_{lu}$ are the collisional rate coefficients (van der Tak et al.~\cite{vandertak07}). The cooling rate $\Lambda$ in erg\,s$^{-1}$\,cm$^{-3}$ is defined as positive for net cooling. In Fig. ~\ref{cooling_fig} the line cooling from o-H$_2$O is compared with that from CO and H$_2$ for the W~Hya and IK~Tau models used in the present analysis. In the region where H$_2$O is abundant its cooling rate will be more than an order of magnitude higher than that from CO. Due to the constraints set by the CO-line intensity ratios, the kinetic temperature profile determined through the CO line modelling is most accurate in the region in which the CO lines are emitted. Any major cooling due to $\rm{H_2O}$ must therefore be partly compensated by additional heating, by adjusting the $h$-parameter so that the kinetic temperature structure is consistent with the observed CO rotational line intensities.

To include molecular cooling from water self-consistently, the energy balance equations for CO and $\rm{H_2O}$ must be solved simultaneously. For a complete description of the energy balance, other processes (such as a complicated heating term) would have to be included as well, which is beyond the scope of this paper. A crude estimate of how $\rm{H_2O}$ line cooling and other processes might affect the modelled line intensities is done by changing the temperature profile resulting from the CO modelling by $\pm\,20\%$.  This changes the line intensities on average by less than $\pm15\%$, without changing the line ratios.

\begin{figure}
   \centering
   \includegraphics[width=8.5cm]{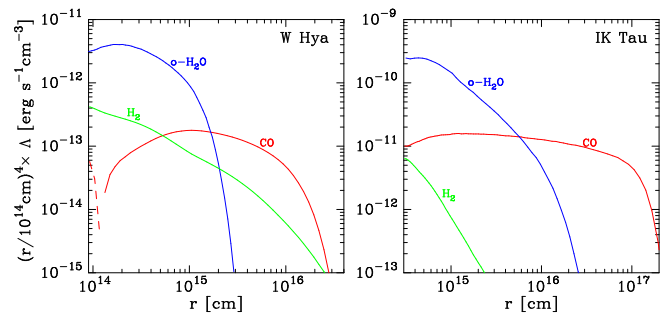}
   \caption{The emissivity $\Lambda$ in erg\,s$^{-1}$\,cm$^{-3}$, scaled by ($r/10^{14}$cm)$^4$, of the line cooling rate from o-H$_2$O, CO and H$_2$ in the circumstellar envelopes around W Hya and IK Tau. There is a net cooling ($\Lambda$\,$>$\,0) throughout the envelopes for these molecules, except very close to the star in the case of CO in W Hya ($\Lambda$\,$<$\,0 indicated by a dashed line). }
              \label{cooling_fig}
    \end{figure}

\subsection{Predictions for HIFI}
\label{hifi}
HIFI (Heterodyne instrument for the far infrared) is a heterodyne receiver onboard the Herschel space telescope, scheduled for launch in 2008. The instrument is a high-resolution (up to $10^7$, $0.03-300\,\rm{km\,s^{-1}}$) heterodyne instrument designed to provide a wide and continuous frequency coverage. The frequency range covered includes a large amount of molecular lines (de Graauw et al.~\cite{degraauwetal}). We have modelled 19 ortho-$\rm{H_2O}$ lines covering a range of excitation and optical depths within the range 557 GHz to 1885 GHz. 

The sensitivity limit of HIFI for a $5\,\sigma$ line detection after 1 hour of integration at a resolution of $R$\,$=$\,$100000$ is between 2.2 and 22.0$\rm{\,\times\,10^{-30}\,W\, cm^{-2}\,Hz^{-1}}$ (de Graauw et al.~\cite{degraauwetal}), depending on the frequency band. The modelled HIFI lines generally have fluxes of $\rm{10^{-27}\,W\,cm^{-2}\,Hz^{-1}}$ or higher, making them observable without difficulty. Typically, between 1 and 10 hours of integration time is required to obtain a sufficient signal-to-noise ratio, even for weaker lines in the first excited vibrational state.

Figure~\ref{hifilines} shows a selection of the modelled lines obtained from the best-fit models for TX~Cam, W~Hya and IK~Tau as given in Table~\ref{modres} (based on the theoretical $\rm{H_2O}$ envelope radius given in Eq.~\ref{outrad}). The dot-dashed line marks the stellar $v_{LSR}$ and the solid line marks $\pm v_{exp}$. Maser action is present in the $5_{32}-4_{41}$ and $6_{34}-5_{41}$ transitions at 621 GHz and 1158 GHz, respectively,  in all three sources. 

The line profiles for models with lower mass-loss rates generally appear to suffer more from self-absorption and show significant emission outside the range $v_{LSR}\pm v_{exp}$. The CSEs of these sources are relatively small and the optical depths at the outer parts of the CSEs are around, or larger than, unity (see Fig.~\ref{ex}). This leads to effective self-absorption on the blue-shifted side, but also to a contribution of the local line profiles to the line wings. This is particularly pronounced in the lines where the optical depths at line centre can reach up to $\sim$100 (Fig.~\ref{ex}) for these kinds of objects, such as $1_{10}-1_{01}$ and  $3_{03}-2_{12}$ (Fig.~\ref{hifilines}). The broadening of the line profiles is further enhanced by the fact that the thermal motion, $v_{th}$, is of the same order or larger when compared with the microturbulent motion, $v_{turb}$, in the emitting region. The result is that optically thin lines can also have significant line broadening (Fig.~\ref{hifiprofiles}). These effects are less significant in the high-mass-loss-rate objects, as the optical depths in the outer parts of these CSEs, which contribute most to the observed emission, are around, or below, one. Also, for the high-mass-loss-rate objects $v_{th} << v_{turb}$ in the outer parts of the envelope.

The sensitivity of the HIFI lines to changes in various parameters has also been tested. A change in abundance by $\pm50\%$ changes the HIFI line fluxes on average by  $\approx\,\pm30\%$ compared to the best-fit model. This compares to the $10-20\%$ change in the ISO lines, indicating that the lines in the range of HIFI are somewhat more sensitive to the $\rm{H_2O}$ abundance (see below). Increasing and decreasing $r_{\rm{e}}$ by a factor of two change the line fluxes on average by less than $\pm16\%$. The line fluxes for the HIFI lines are more sensitive for a -20\% change of the expansion velocity than the ISO lines (+20\% on average; however, optically thin lines are more sensitive, see below). Resolved line profiles will be crucial for getting information on the dynamical structure of the inner CSE.

\begin{table}
\caption{Model maximum intensities for 3 optically thin lines within the range of HIFI (in units of $10^{-30}\,\rm{W\,cm^{-2}\,Hz^{-1}}$).}
\label{thin}
\centering
\begin{tabular}{c c c c}
\hline\hline
Trans 			& $\nu$Ê	& R~Dor 	& TX~Cam\\
$J_{K_-K_+}-J_{K_-K_+}$& [GHz] & & \\
\hline
$8_{54}-7_{61}$	& 1169	& 10		& 4Ê\\
$8_{54}-9_{27}$	& 1596	& 3		& 1 \\
$8_{45}-7_{52}$	& 1885	& 320	& 40 \\
\hline
\end{tabular}
\end{table}

Of the 19 modelled lines, 5 are optically thin throughout the CSE. Two of these are maser lines (the $5_{32}-4_{41}$ and $6_{34}-5_{41}$ transitions at 621 GHz and 1158 GHz, respectively). The remaining three lines are shown in Table~\ref{thin} with their predicted maximum intensities. The $5\,\sigma$ line detection limit after 1 hour of integration time at $R=100000$ is $\approx2\,\times\,\rm{10^{-29}\,W\,cm^{-2}\,Hz^{-1}}$, making it difficult to detect some of these lines.

The sensitivity of these lines to the ortho-$\rm{H_2O}$ abundance, the outer radius, mass-loss rate, kinetic temperature profile, and expansion velocity was tested for R~Dor and TX~Cam. The resulting change in line fluxes is presented in Tables~\ref{hifitest_rdor} and~\ref{hifitest_txcam} for R~Dor and TX~Cam, respectively. Except for the $8_{54}-9_{27}$ transition, the optically thin lines are much more sensitive to the abundance, mass-loss rate and expansion velocity than the ISO lines are. The P-Cygni profile causes the insensitivity of the integrated flux in the $8_{54}-9_{27}$ transition. The high sensitivity to the expansion velocity is due to the change in the density profile as a result of keeping the mass-loss rate constant while decreasing the expansion velocity. They are only slightly more sensitive to a change in the kinetic temperature profile compared to the ISO lines, and not at all sensitive to the outer radius.  
For R~Dor, transitions $3_{21}-3_{12}$, $2_{12}-1_{01}$, and $2_{21}-2_{12}$ at 1163 GHz, 1670 GHz, and 1661 GHz, respectively, are more sensitive to the change of the outer radius (with up to a 30\% change in line flux), but the emission region is essentially excitation limited.

\begin{table}
\caption{Parameter dependence of the model fluxes of three optically thin lines observable by HIFI for the low-mass-loss-rate object R~Dor. Column 2 gives the change of the various parameters relative to their best-fit value (Table~\ref{modres}). The remaining columns give the change of the model fluxes compared to the best-fit fluxes for the respective lines.}
\label{hifitest_rdor}
\centering
\begin{tabular}{l c c c c}
\hline\hline
param. & change &$8_{54}-7_{61}$ & $8_{54}-9_{27}$& $8_{45}-7_{52}$  \\
\hline
$f_{\rm{e}}$     & {\phantom{1}}$\pm50\%$& $\pm30\%$ & {\phantom{1}}$\pm6\%$ & {\phantom{1}}$\pm55\%$\\
$R_{\rm{mod}}$	& +100\%              & {\phantom{1}}+4\% & {\phantom{1}}+5\% & {\phantom{1}}{\phantom{1}}+2\%\\
				& {\phantom{1}}--50\% &{\phantom{1}}--2\% &{\phantom{1}} --2\% & {\phantom{1}}{\phantom{1}}--4\%\\
$\emph{\.M}$		& {\phantom{1}}+50\%&+77\% & +10\% & +153\%\\
				& {\phantom{1}}--50\% & --25\% & {\phantom{1}}--7\% & {\phantom{1}}--45\%\\
$v_{\rm{exp}}$	& {\phantom{1}}--20\% & +28\% & {\phantom{1}}+5\%&{\phantom{1}}+55\%	\\
$T_{\rm{kin}}$	& {\phantom{1}}+20\% &+21\% & {\phantom{1}}+1\% & {\phantom{1}}+55\%\\
			& {\phantom{1}}--20\% & {\phantom{1}}+9\% & {\phantom{1}}--1\% & {\phantom{1}}+20\% \\
\hline
\end{tabular}
\end{table}

\begin{table}
\caption{Same as Table~\ref{hifitest_rdor}, but for the intermediate-mass-loss-rate object TX~Cam.}
\label{hifitest_txcam}
\centering
\begin{tabular}{l c c c c}
\hline\hline
param. & change &$8_{54}-7_{61}$ & $8_{54}-9_{27}$& $8_{45}-7_{52}$  \\
\hline
$f_{\rm{e}}$     & {\phantom{1}}$\pm50\%$& $\pm30\%$ & {\phantom{1}}$\pm7\%$ & {\phantom{1}}$\pm30\%$\\
$R_{\rm{mod}}$	& +100\%              & +20\% & {\phantom{1}}+3\% & {\phantom{1}}{\phantom{1}}+6\%\\
				& {\phantom{1}}--50\% &--14\% &{\phantom{1}} --2\% & {\phantom{1}}{\phantom{1}}--6\%\\
$\emph{\.M}$		& {\phantom{1}}+50\%&+70\% & +14\% & {\phantom{1}}+71\%\\
				& {\phantom{1}}--50\% & --44\% & --11\% & {\phantom{1}}--53\%\\
$v_{\rm{exp}}$	& {\phantom{1}}--20\% & +48\% & {\phantom{1}}+8\%&{\phantom{1}}+40\%	\\
$T_{\rm{kin}}$	& {\phantom{1}}+20\% &+13\% & {\phantom{1}}+2\% & {\phantom{1}}+22\%\\
			& {\phantom{1}}+20\% &+13\% & {\phantom{1}}+2\% & {\phantom{1}}+22\%\\
\hline
\end{tabular}
\end{table}

An example of how the  $8_{45}-7_{52}$ line profile changes between the different models for R~Dor (top) and TX~Cam (bottom) is shown in Fig.~\ref{hifiprofiles}. The left panels show models in which the parameters were reduced as indicated in Tables~\ref{hifitest_rdor} and~\ref{hifitest_txcam}, and the right panels show the models with increased parameters. The line profiles change mainly in intensity. However, for TX~Cam the optical depth in the $8_{45}-7_{52}$ transition is just below one in the best fit model. The line therefore changes from optically thick to optically thin for the different models, and as the optical depth decreases, the line becomes increasingly flat-topped. The detailed abundance distribution is likely to have the largest effect on the shape of the line profile. Detailed examination of the predicted HIFI line profiles based on the spectrally resolved 557 GHz lines of R~Cas, R~Dor and W~Hya, obtained with the Odin satellite, will be done in a forthcoming paper (Maercker et al., in prep.). Since this line is excited throughout the CSE, it will in particular help put better constraints on the size of the $\rm{H_2O}$ emitting region.

 \begin{figure}
   \centering
   \includegraphics[width=9cm]{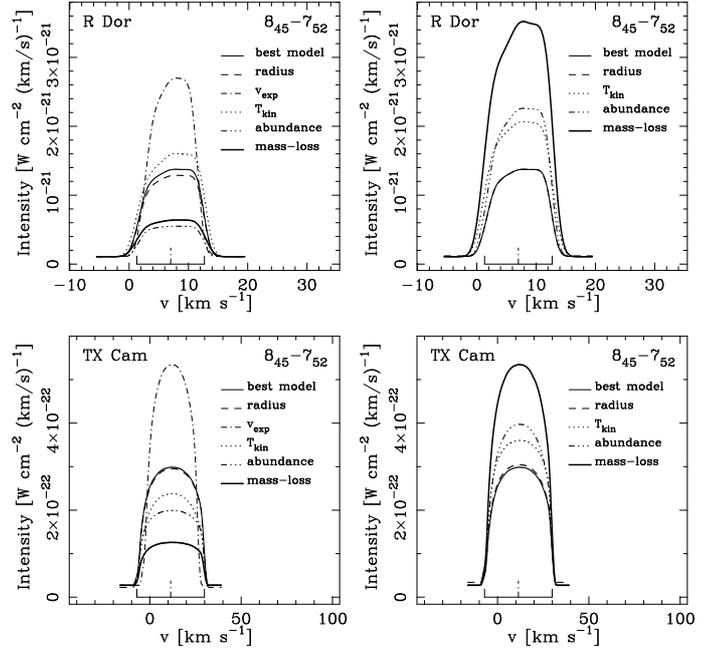}
   \caption{Line profiles of the $8_{45}-7_{52}$ transition for the various models indicated in Tables~\ref{hifitest_rdor} and~\ref{hifitest_txcam} for R~Dor (top) and TX~Cam (bottom). The left panels show the models with decreased parameters; the right panels show models with increased parameters.One parameter is varied, while the others are held fixed, thus leading to a change in the density profile when decreasing the expansion velocity.}
              \label{hifiprofiles}
    \end{figure}

  \begin{figure*}
   \centering
   \includegraphics{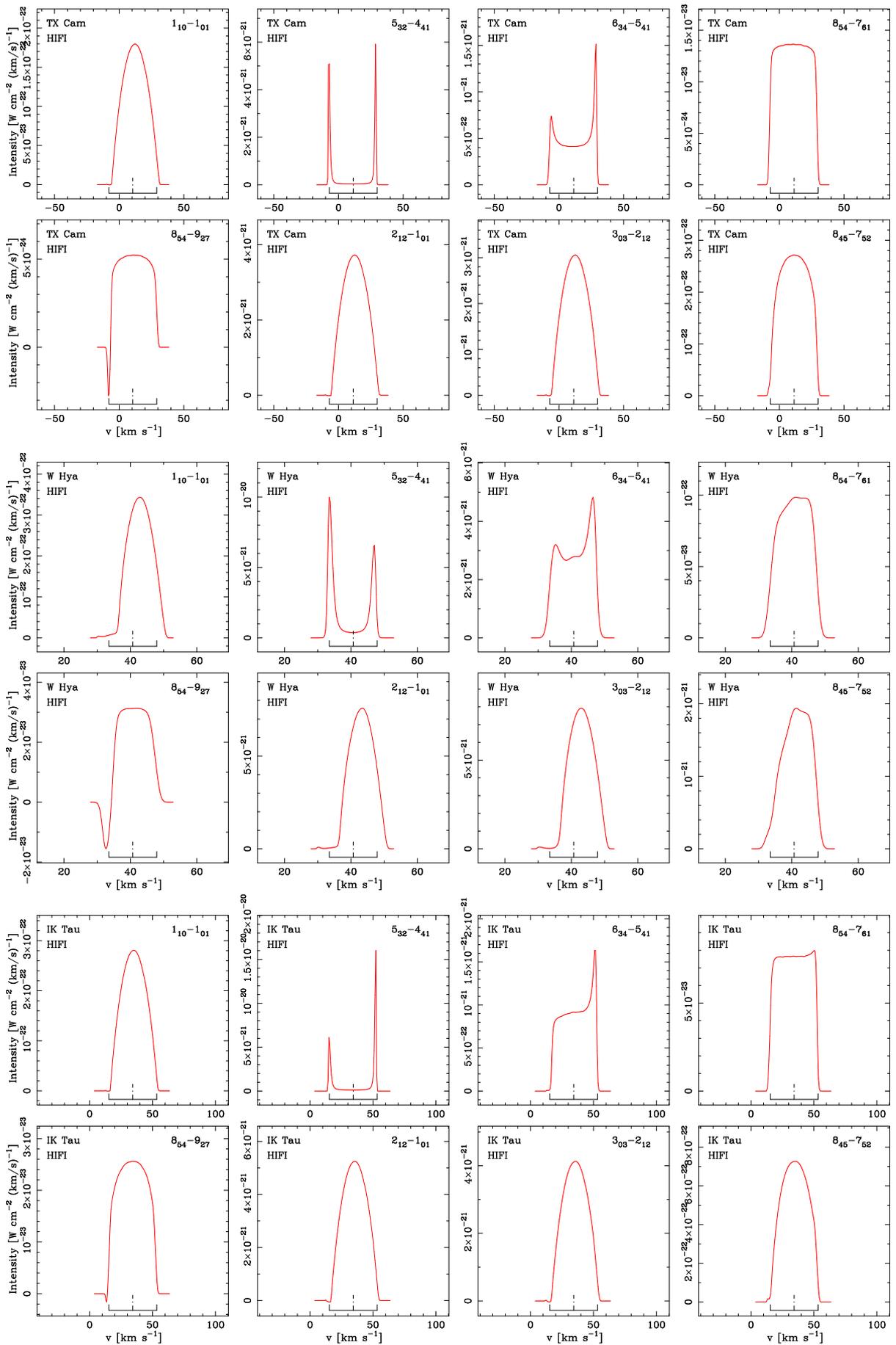}
   \caption{Eight ortho-H$_2$O lines that can be observed with Herschel/HIFI  TX~Cam (top two rows),  W~Hya (middle two rows) and for IK~Tau (bottom two rows) models based on the theoretical radius from Eq.~\ref{outrad}. The dot-dashed line marks the stellar $v_{\rm{LSR}}$, the solid line shows  $v_{\rm{LSR}}\pm v_{\rm{exp}}$. A number of line profile shapes are predicted, and the HIFI observations will help to verify or adjust the models.}
              \label{hifilines}
    \end{figure*}

\subsection{ISO CO-lines}
\label{isoco}
The ISO LWS spectrum also contains high-lying transitions of the CO molecule (Fig.~\ref{isolines}). These CO lines probe the same region as the $\rm{H_2O}$ lines, while the radio lines probe the regions further out. Modelling these high excitation CO  lines can be used as a consistency check for the adopted circumstellar model used in the ALI models.  Due to weak lines and blending, in general only between one and three lines were found in the ISO spectra. W~Hya, R~Dor, and R~Cas show the CO $J$\,$=$\,16\,$\rightarrow$\,15 and $J$\,$=$\,22\,$\rightarrow$\,21 lines. R~Cas also shows the CO $J$\,$=$\,17\,$\rightarrow$\,16 line. In the remaining objects only the CO $J$\,$=$\,16\,$\rightarrow$\,15 line could be found. For all objects except W~Hya, the best-fit model fits the observations within $\pm20\%$. The modelled CO lines for W~Hya are $\sim$\,80\% weaker than the observations, in line with the higher mass-loss rate used by Justtanont et al. (\cite{justtanontetal}) (see Sect.~\ref{COlinemod}). 

\section{Discussion}
\label{disc}

\subsection{Comparison with previous results}
Abundance estimates for two of the sources treated here have been made previously. For W Hya, Justtanont et al. (\cite{justtanontetal}) find an ortho-$\rm{H_2O}$ abundance of 1.0\,$\pm$\,0.4\,$\times10^{-3}$ relative to $\rm{H_2}$. In their paper they use an ALI code to determine the $\rm{H_2O}$ abundance by modelling of ISO SWS and LWS lines, and the Odin 557 GHz line. This agrees well with the abundance determined here (1.5\,$\pm$\,0.75\,$\times$\,$10^{-3}$). Our higher abundance is due to the lower mass-loss rate used (see Sect.~\ref{COlinemod}). Barlow et al. (\cite{barlowetal}) employ a large velocity gradient (LVG) model and determine a total $\rm{H_2O/H_2}$ of 8\,$\times$\,$10^{-4}$ in the inner part of the CSE ($r$\,$\leq$\,6\,$\times$\,$10^{14}$\,cm) and 3\,$\times$\,$10^{-4}$ in the outer part ($r$\,$>$\,6\,$\times$\,$10^{14}$\,cm). Although the ortho-to-para ratio in their article is not very well constrained, between 1 and 3, their results are consistent with the abundances derived here. Zubko \& Elitzur (\cite{zubkoelitzur}) determine an abundance for ortho-$\rm{H_2O}$ of 5.7\,$\times$\,$10^{-5}$, one order of magnitude lower than the results presented here, and a mass-loss rate one order of magnitude higher (2.3\,$\times$\,$10^{-6}$\,M$_{\odot}$\,yr$^{-1}$). Their results are therefore consistent with ours, given the higher mass-loss rate. For R~Cas, Truong-Bach et al. (\cite{truongbachetal}) determine an ortho-$\rm{H_2O}$ abundance of only 7.9\,$\times$\,$10^{-6}$, two orders of magnitude lower than our result, using ISO LWS spectra. The mass-loss-rate used in their paper is somewhat lower than the one determined from the CO modelling here, further increasing the discrepancy in the abundance estimates. They do not model the dust emission explicitly, but instead use a power law to reproduce the far-infrared continuum spectrum. They also estimate cooling by CO emission by adopting a three level model, compared to the more complex scheme included here. On the other hand, they include a dust grain drift velocity (relative to the gas) that depends on the distance to the star. Including a velocity field in our models would probably result in somewhat lower abundances. Although the models are different, this is far from enough to explain the different results.

\subsection{Evaporating planetary systems?}
\label{planet}

Assuming an ortho-to-para ratio of 1 gives an estimate of the total $\rm{H_2O}$ abundance of $\sim$\,6\,$\times$\,$10^{-4}$ in the majority of our stars. [Justtanont et al. (\cite{justtanontetal}) and Zubko \& Elitzur (\cite{zubkoelitzur}) found an ortho-to-para $\rm{H_2O}$ ratio of 1 and 0.8 for W~Hya, respectively. Truong-Bach et al. (\cite{truongbachetal}) derive a ratio of 3 for R~Cas, in accordance with what is expected in LTE.] The estimated W~Hya $\rm{H_2O}$ abundance is significantly higher than this, but, as outlined in Justtanont et al. (\cite{justtanontetal}), the W~Hya mass-loss rate could be significantly higher than the value we use, and consequently the $\rm{H_2O}$ abundance could be significantly lower. The cosmic abundance of carbon and oxygen (Anders \& Grevesse~\cite{andersgrevesse}) limits the abundance of $\rm{H_2O}$ relative to $\rm{H_2}$ to not more than $1\,\times\,10^{-3}$, assuming that all the carbon goes into CO, and all the remaining O goes into $\rm{H_2O}$. The abundance of $\rm{H_2O}$ in thermal equilibrium (TE) is approximately  1-3$\times\,10^{-4}$ relative to $\rm{H_2}$ (Mamon et al.~\cite{mamonetal87}; Nejad \& Miller~\cite{nejadco88}; Willacy \& Millar~\cite{willacyco97}; Duari et al.~\cite{duarietal99}). Taken at face value, our results indicate that the abundances of $\rm{H_2O}$ observed in M-type AGB stars cannot be explained by TE chemistry. The uncertainty in our results is, however, at least a factor of a few, and therefore does not rule out a TE abundance. NLTE chemistry can affect the $\rm{H_2O}$ abundance, e.g., due to different densities and temperatures in shocked regions close to the star (Duari et al.~\cite{duarietal99}; Cherchneff~\cite{cherchneff}), and photodissociation further out in the envelope will strongly affect the abundance (Mamon et al.~\cite{mamonetal87}; Nejad \& Miller~\cite{nejadco88}; Willacy \& Millar~\cite{willacyco97}). Other processes suggested to explain the relatively high water abundances are the evaporation of icy bodies (Justtanont et al.~\cite{justtanontetal}) or grain surface reactions, such as Fischer-Tropsch catalysis on the surfaces of small metallic grains (Willacy~\cite{willacy}). The former process was first suggested to explain the surprising detection of $\rm{H_2O}$ in C-rich CSEs (Melnick et al.~\cite{melnicketal01}; Hasegawa et al.~\cite{hasegawaetal}).

Future observations with Herschel will make it possible to discriminate between the different formation scenarios, as suggested by e.g. Gonz\'alez-Alfonso et al.~(\cite{gonzalezetal2007}) in the case of C-rich CSEs. A high abundance of $\rm{H_2O}$ in the inner regions of the CSE will be observable through a significant amount of mid-excitation $\rm{H_2O}$ transitions. These are already observed in the ISO spectra for our objects, indicating that at least some of the observed $\rm{H_2O}$ is formed in the inner envelope. Both Fischer-Tropsch catalysis and the evaporation of cometary bodies result in similar shell-like $\rm{H_2O}$ distributions. The evaporation hypothesis, however, implies the presence of HDO, the detection of which would rule out the formation of $\rm{H_2O}$ from material ejected by the star.

\subsection{The very low abundance in WX~Psc}
\label{wxpsc}
WX~Psc has an estimated abundance that is two orders of magnitude lower than for the other objects. A possible explanation for this is depletion of $\rm{H_2O}$ on dust grains. For a sample of carbon and M-type stars, Sch\"oier \& Olofsson (\cite{schoierolofsson2006}) show that the abundance of SiO decreases in the CSEs with increasing \emph{\.M}$/v_{\rm{exp}}$ (a measure of the density), and explain this by depletion of SiO onto the dust grains.  The freezing of water ice onto dust grains is also the explanation for the lack of $\rm{H_2O}$ lines in the spectra of a sample of OH/IR stars (Justtanont et al.~\cite{justtanontetal06}). Another possible explanation for the low abundance in WX~Psc could be that the ISO water vapour observations are probing a present day mass-loss rate, whereas the CO transitions probe an earlier mass-loss epoch (Ramstedt et al.~\cite{ramstedtetal07}). If the present day mass loss is much lower, this would require an increase in the $\rm{H_2O}$ abundance, thereby putting WX~Psc  more in line with the other objects. Mass-loss-rate modulations in  WX~Psc with a low present day mass-loss-rate have been suggested previously by Kemper et al. ({\cite{kemperetal}) and are further discussed by Ramstedt et al. (\cite{ramstedtetal07}).

 \section{Conclusions}
 \label{conc}
We have demonstrated the ability of our accelerated lambda iteration (ALI) code to model circumstellar $\rm{H_2O}$ emission lines in M-type AGB stars at optical depths $>$\,1000, and ortho-$\rm{H_2O}$ abundances are estimated. The models include the ground and first excited vibrational states, collisions within the ground vibrational state, the first excited vibrational state, and between these states, and excitation by the stellar and dust radiation fields. The sample of stars modelled spans one order of magnitude in mass-loss rate. The ortho-$\rm{H_2O}$ abundances determined, in the range (0.02--15)$\times$10$^{-4}$, are estimated to be accurate to within 50\%, within the adopted circumstellar model, while the absolute error (including model-dependent assumptions) is within a factor of 5. The estimated $\rm{H_2O}$ abundances are high in comparison to what can be produced by chemical models, but we explain that the estimates are rather uncertain. 

The outer radius of the $\rm{H_2O}$ envelope is better determined for the low-mass-loss-rate objects and corresponds to within 1$\sigma$ to the radius given by the theoretical estimate in Eq.~\ref{outrad}.

Including excitation of the ground state through the vibrationally excited state affects the low-mass-loss-rate objects, increasing the model line fluxes by $\approx\,40\%$. This results in a lower abundance for these objects compared to models that do not include the excited vibrational state (Justtanont et al.~\cite{justtanontetal}). For higher mass-loss-rate objects the vibrational excitation is less important.

Several formation processes for the observed $\rm{H_2O}$ abundances are possible. LTE or NLTE models may explain the formation of $\rm{H_2O}$ in the innermost layers of the CSE, whereas Fischer-Tropsch catalysis or the evaporation of Kuiper-belt like objects are possible sources of $\rm{H_2O}$ in the outer parts of the envelope. Future observations with the Herschel telescope will help to differentiate between these scenarios. The HIFI lines are generally less saturated and will therefor give a better estimate of the circumstellar $\rm{H_2O}$ abundance. In particular, optically thin lines, such as the $8_{45}-7_{52}$ transition, are sensitive to the $\rm{H_2O}$ abundance, and will help constrain this parameter. The ground state line at 557 GHz, on the other hand, is excited throughout the envelope and will help in setting constraints on the size of the $\rm{H_2O}$ envelope.

For WX~Psc, we suggest that sticking of $\rm{H_2O}$ onto dust grains may explain the observed low amount of $\rm{H_2O}$. Alternatively, the present day mass-loss rate is low.

\begin{acknowledgements}
The authors would like to thank the anonymous referee for useful comments that helped improve the quality of the manuscript.
The authors acknowledge the financial support from the Swedish Research Council and the Swedish National Space Board.
\end{acknowledgements}

\newpage
\appendix
\section{Accelerated lambda iteration (ALI)}
\label{ali}

\subsection{General considerations}
The radiative transfer models used in this work are created using the accelerated lambda iteration (ALI) method (Rybicki \& Hummer~\cite{rybickihummer91}, \cite{rybickihummer92}). As for most NLTE codes, the ALI technique makes an initial guess of the level populations $n_i$ and solves the statistical equilibrium (SE) equations (for transition $l \to l'$)
\begin{eqnarray}
\label{SE}
\sum_{l'<l}\lbrack n_lA_{ll'}-(n_{l'}B_{l'l}-n_lB_{ll'})\bar{J}_{ll'}\rbrack -\nonumber\\
\sum_{l'>l}\lbrack n_{l'}A_{l'l}-(n_{l}B_{ll'}-n_{l'}B_{l'l})\bar{J}_{ll'}\rbrack+\nonumber\\
\sum_{l'}(n_lC_{ll'}-n_{l'}C_{l'l})=0,
\end{eqnarray}
where $A_{ll'}$ is the Einstein $A$-coefficient, $B_{ll'}$ and $B_{l'l}$ are the Einstein $B$-coefficients for stimulated emission and absorption respectively, and $C_{ll'}$ are the collisional rates. $\bar{J}_{ll'}$ is the mean integrated intensity
\begin{equation}
\label{meanint}
\bar{J}_{ll'}=  {1\over 4\pi} \int{d\Omega}\int{d\nu \phi_{ll'}(\mu,\nu)I(\mu,\nu)},
\end{equation}
where $I(\mu,\nu)$ is the specific intensity at frequencey $\nu$ along direction $\mu$, and $\phi_{ll'}(\mu,\nu)$ is the normalized line shape profile (Doppler profile) used as a weight function. 
In the spherically symmetric case it is enough to solve the SE equations by determining $\bar{J}_{ll'}$ at every radial point in the circumstellar shell, considering both the local and non-local contributions from all directions $\mu$. This is done by introducing the lambda operator $\Lambda_{\mu\nu}$, which relates to the specific intensity $I(\mu,\nu)$ as
\begin{equation}
\label{speint}
I(\mu,\nu)=\Lambda_{\mu\nu}S\rm{_{tot}}(\mu,\nu)+L_{bg},
\end{equation}
where $S\rm{_{tot}}(\mu,\nu)$ is the total source function (including dust and all overlapping lines) and $L_{bg}$ is the contribution from the background to $I(\mu,\nu)$. The lambda operator can be seen as a $N \times N$ matrix with $N$ radial points. The diagonal elements of the matrix describe the local contribution to the specific intensity, while the off-diagonal elements describe the contribution from all other parts of the cloud. If $\tau(\mu,\nu)$ is the local optical depth in direction $\mu$, the diagonal elements become
\begin{equation}
\label{diag}
\Lambda_{r,r}(\mu,\nu)=1-e^{-\tau(\mu,\nu)}.
\end{equation}
For high optical depths ($\tau(\mu,\nu)\gg1$) $\Lambda_{r,r}\to 1$, whereas in the optical thin case ($\tau(\mu,\nu)\to0$) $\Lambda_{r,r}\to 0$. Instead of applying the lambda iteration directly, the accelerated lambda iteration technique uses an approximate lambda operator $\Lambda_{\mu\nu}^*$ to take care of the high optical depths separately
\begin{equation}
\label{approx}
I(\mu,\nu)=\Lambda_{\mu\nu}^*S_{tot}(\mu,\nu)+(\Lambda_{\mu\nu}-\Lambda_{\mu\nu}^*)S_{tot}^{\dag}(\mu,\nu).
\end{equation}
$S\rm{_{tot}}^{\dag}(\mu,\nu)$ is the source function from the previous iteration and can be used, since convergence guarantees that  $S\rm{_{tot}}^{\dag}\to S\rm{_{tot}}$. It is convenient to simply choose the diagonal elements of the lambda operator as the approximate lambda operator. This makes the matrix inversion easily possible, and the SE equations can be solved at every radial point individually, based on the approximate lambda operator only. Using the accelerated lambda iteration technique improves the rate of convergence compared to normal lambda iteration, particularly in cases of high optical depths. To further improve convergence, Ng acceleration (Ng~\cite{ng}) is used, with a good description of the method given by Olson et al. (\cite{olsonetal}).

\begin{figure}
   \centering
   \includegraphics[width=8.5cm]{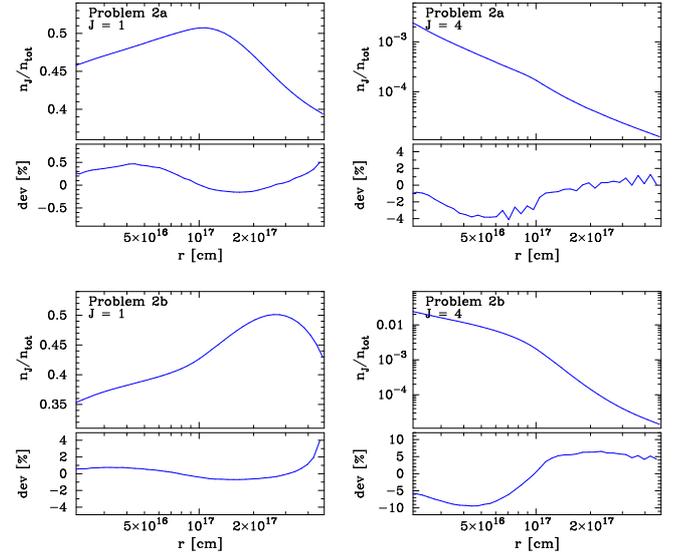}
   \caption{Results from benchmarking our ALI code against the results found in van Zadelhoff et al.\ (\cite{zadelhoffetal}) for their problems 2a and 2b, consisting of an inside-out collapsing spherical cloud. The figure shows the radial distributions of the relative number density of HCO$^{+}$ molecules in the rotational levels $J$\,$=$\,1 and $J$\,$=$\,4 as obtained by our ALI code after convergence. Also shown are the deviations (in percent) of our results when compared to the mean of the results found by all other radiative transfer codes, as presented in van Zadelhoff et al.\ (\cite{zadelhoffetal}).}
              \label{problem2}
    \end{figure}
    
\subsection{Benchmarking}
When developing a new radiative transfer code it is important to perform detailed tests as to its accuracy. A number of such test problems are available for download at {\tt www.strw.leidenuniv.nl/$\sim$radtrans}. Our ALI code has been benchmarked against the solutions to these problems as found by seven different radiative transfer codes (presented in van Zadelhoff et al.~\cite{zadelhoffetal}). We find that our ALI code reproduces the results for all these test problems with an accuracy that matches all the other codes included in van Zadelhoff et al.\ (\cite{zadelhoffetal}).

As an example, Fig.~\ref{problem2}  shows the solution obtained from our ALI code for problems 2a and 2b of van Zadelhoff et al.\ (\cite{zadelhoffetal}). These problems consist of an inside-out collapsing spherical cloud and contain radial gradients in most physical parameters such as velocity, density, temperature and micro-turbulent velocity. The excitation analysis is performed for HCO$^{+}$, and the final level populations (in relation to the total number density of HCO$^{+}$ molecules, $n_{\mathrm{tot}}$) for the $J$\,$=$\,1 and $J$\,$=$\,4 rotational levels are shown in Fig.~\ref{problem2}. Also shown are the deviations of our ALI code (in percent) from the mean of the results obtained by the other radiative transfer codes presented in van Zadelhoff et al.\ (\cite{zadelhoffetal}). The spread in the results, around the mean, between various codes in van Zadelhoff et al.\ (\cite{zadelhoffetal}) is for the $J$\,$=$\,1 level  about $\pm$\,1\% and $\pm$\,2\% for problems 2a and 2b, respectively. For the $J$\,$=$\,4 level a larger spread of about $\pm$\,5\% and $\pm$\,15\% was obtained for problems 2a and 2b, respectively.

\end{document}